\documentclass[letterpaper,twocolumn,superscriptaddress,groupedaddress,titlepage,nofootinbib,amsmath]{revtex4}

\usepackage{bm,color}
\usepackage{graphicx}
\usepackage{amssymb, amsmath}
\usepackage[amssymb]{SIunits}
\usepackage{mathtools}
\usepackage{multirow}
\usepackage{slashbox}

\def\camb{{\sc CAMB}}
\def\herschelspire{{\it Herschel}/SPIRE}

\newcommand{\Pb}{\textsc{Polarbear}}
\newcommand{\pb}{\textsc{Polarbear}}

\newcommand{\lcdm}{$\Lambda$CDM}

\newcommand{\herschel}{\textit{Herschel}}

\newcommand{\apjl}{Astrophysical Journal Letters}
\newcommand{\apjs}{Astrophysical Journal Supplement Series}
\newcommand{\aap}{Astronomy and Astrophysics}

\begin{document}

\title{Measurement of the Cosmic Microwave Background Polarization Lensing Power
Spectrum with the POLARBEAR experiment}

\author{\pb\ Collaboration} \noaffiliation
\author{P.A.R. Ade} \affiliation{School of Physics and Astronomy, Cardiff University}
\author{Y. Akiba} \affiliation{The Graduate University for Advanced Studies}
\author{A.E. Anthony} \affiliation{Center for Astrophysics and Space Astronomy, University of Colorado, Boulder}
\author{K. Arnold} \affiliation{Department of Physics, University of California, San Diego}
\author{M. Atlas} \affiliation{Department of Physics, University of California, San Diego}
\author{D. Barron} \affiliation{Department of Physics, University of California, San Diego}
\author{D. Boettger} \affiliation{Department of Physics, University of California, San Diego}
\author{J. Borrill} \affiliation{Computational Cosmology Center, Lawrence Berkeley National Laboratory} \affiliation{Space Sciences Laboratory, University of California, Berkeley}
\author{S. Chapman} \affiliation{Department of Physics and Atmospheric Science, Dalhousie University, Halifax, NS, B3H 4R2, Canada}
\author{Y. Chinone} \affiliation{High Energy Accelerator Research Organization (KEK)} \affiliation{Department of Physics, University of California, Berkeley}
\author{M. Dobbs} \affiliation{Physics Department, McGill University}
\author{T. Elleflot} \affiliation{Department of Physics, University of California, San Diego}
\author{J. Errard} \affiliation{Computational Cosmology Center, Lawrence Berkeley National Laboratory}\affiliation{Space Sciences Laboratory, University of California, Berkeley}
\author{G. Fabbian} \affiliation{AstroParticule et Cosmologie, Univ Paris Diderot, CNRS/IN2P3, CEA/Irfu, Obs de Paris, Sorbonne Paris Cit\'e, France} \affiliation{International School for Advanced Studies (SISSA)}
\author{C. Feng {\footnote{Corresponding author: \href{mailto: cfeng@physics.ucsd.edu}{cfeng@physics.ucsd.edu}}}}\affiliation{Department of Physics, University of California, San Diego}
\author{D. Flanigan} \affiliation{Department of Physics, University of California, Berkeley} \affiliation{Columbia University}
\author{A. Gilbert} \affiliation{Physics Department, McGill University}
\author{W. Grainger} \affiliation{Rutherford Appleton Laboratory, STFC}
\author{N.W. Halverson} \affiliation{Center for Astrophysics and Space Astronomy, University of Colorado, Boulder} \affiliation{Department of Astrophysical and Planetary Sciences, University of Colorado, Boulder} \affiliation{Department of Physics, University of Colorado, Boulder}
\author{M. Hasegawa} \affiliation{High Energy Accelerator Research Organization (KEK)} \affiliation{The Graduate University for Advanced Studies}
\author{K. Hattori} \affiliation{High Energy Accelerator Research Organization (KEK)}
\author{M. Hazumi} \affiliation{High Energy Accelerator Research Organization (KEK)} \affiliation{The Graduate University for Advanced Studies} \affiliation{Kavli Institute for the Physics and Mathematics of the Universe (WPI), Todai Institutes for Advanced Study, The University of Tokyo}
\author{W.L. Holzapfel} \affiliation{Department of Physics, University of California, Berkeley}
\author{Y. Hori} \affiliation{High Energy Accelerator Research Organization (KEK)}
\author{J. Howard} \affiliation{Department of Physics, University of California, Berkeley} \affiliation{University of Oxford}
\author{P. Hyland} \affiliation{Physics Department, Austin College}
\author{Y. Inoue} \affiliation{The Graduate University for Advanced Studies}
\author{G.C. Jaehnig} \affiliation{Center for Astrophysics and Space Astronomy, University of Colorado, Boulder} \affiliation{Department of Physics, University of Colorado, Boulder}
\author{A. Jaffe} \affiliation{Department of Physics, Imperial College London}
\author{B. Keating} \affiliation{Department of Physics, University of California, San Diego}
\author{Z. Kermish} \affiliation{Physics Department, Princeton University}
\author{R. Keskitalo} \affiliation{Computational Cosmology Center, Lawrence Berkeley National Laboratory}
\author{T. Kisner} \affiliation{Computational Cosmology Center, Lawrence Berkeley National Laboratory} \affiliation{Space Sciences Laboratory, University of California, Berkeley}
\author{M. Le Jeune} \affiliation{AstroParticule et Cosmologie, Univ Paris Diderot, CNRS/IN2P3, CEA/Irfu, Obs de Paris, Sorbonne Paris Cit\'e, France}
\author{A.T. Lee} \affiliation{Department of Physics, University of California, Berkeley}\affiliation{Physics Division, Lawrence Berkeley National Laboratory}
\author{E. Linder} \affiliation{Space Sciences Laboratory, University of California, Berkeley}\affiliation{Physics Division, Lawrence Berkeley National Laboratory}
\author{E. M. Leitch} \affiliation{Department of Astronomy and Astrophysics, University of
Chicago, 5640 South Ellis Avenue, Chicago, IL 60637, USA}\affiliation{Kavli Institute for Cosmological Physics, University of
Chicago, 5640 South Ellis Avenue, Chicago, IL 60637, USA}
\author{M. Lungu} \affiliation{Department of Physics, University of California, Berkeley}
\author{F. Matsuda} \affiliation{Department of Physics, University of California, San Diego}
\author{T. Matsumura} \affiliation{High Energy Accelerator Research Organization (KEK)}
\author{X. Meng} \affiliation{Department of Physics, University of California, Berkeley}
\author{N.J. Miller} \affiliation{Observational Cosmology Laboratory, Code 665, NASA Goddard Space Flight Center}
\author{H. Morii} \affiliation{High Energy Accelerator Research Organization (KEK)}
\author{S. Moyerman} \affiliation{Department of Physics, University of California, San Diego}
\author{M.J. Myers} \affiliation{Department of Physics, University of California, Berkeley}
\author{M. Navaroli} \affiliation{Department of Physics, University of California, San Diego}
\author{H. Nishino} \affiliation{Kavli Institute for the Physics and Mathematics of the Universe (WPI), Todai Institutes for Advanced Study, The University of Tokyo}
\author{H. Paar} \affiliation{Department of Physics, University of California, San Diego}
\author{J. Peloton} \affiliation{AstroParticule et Cosmologie, Univ Paris Diderot, CNRS/IN2P3, CEA/Irfu, Obs de Paris, Sorbonne Paris Cit\'e, France}
\author{E. Quealy} \affiliation{Department of Physics, University of California, Berkeley} \affiliation{Physics Department, Napa Valley College}
\author{G. Rebeiz} \affiliation{Department of Electrical and Computer Engineering, University of California, San Diego}
\author{C.L. Reichardt} \affiliation{Department of Physics, University of California, Berkeley}
\author{P.L. Richards} \affiliation{Department of Physics, University of California, Berkeley}
\author{C. Ross} \affiliation{Department of Physics and Atmospheric Science, Dalhousie University, Halifax, NS, B3H 4R2, Canada}
\author{I. Schanning} \affiliation{Department of Physics, University of California, San Diego}
\author{D.E. Schenck} \affiliation{Center for Astrophysics and Space Astronomy, University of Colorado, Boulder} \affiliation{Department of Astrophysical and Planetary Sciences, University of Colorado, Boulder}
\author{B. Sherwin} \affiliation{Department of Physics, University of California, Berkeley}\affiliation{Miller Institute for Basic Research in Science, University of California, Berkeley}
\author{A. Shimizu} \affiliation{The Graduate University for Advanced Studies}
\author{C. Shimmin} \affiliation{Department of Physics, University of California, Berkeley}
\author{M. Shimon} \affiliation{Department of Physics, University of California, San Diego}\affiliation{School of Physics and Astronomy, Tel Aviv University, Tel Aviv, Israel} 
\author{P. Siritanasak} \affiliation{Department of Physics, University of California, San Diego}
\author{G. Smecher} \affiliation{Three-Speed Logic, Inc.}
\author{H. Spieler} \affiliation{Physics Division, Lawrence Berkeley National Laboratory}
\author{N. Stebor} \affiliation{Department of Physics, University of California, San Diego}
\author{B. Steinbach} \affiliation{Department of Physics, University of California, Berkeley}
\author{R. Stompor} \affiliation{AstroParticule et Cosmologie, Univ Paris Diderot, CNRS/IN2P3, CEA/Irfu, Obs de Paris, Sorbonne Paris Cit\'e, France}
\author{A. Suzuki} \affiliation{Department of Physics, University of California, Berkeley}
\author{S. Takakura} \affiliation{Osaka University} \affiliation{High Energy Accelerator Research Organization (KEK)}
\author{T. Tomaru} \affiliation{High Energy Accelerator Research Organization (KEK)}
\author{B. Wilson} \affiliation{Department of Physics, University of California, San Diego}
\author{A. Yadav} \affiliation{Department of Physics, University of California, San Diego}
\author{O. Zahn} \affiliation{Physics Division, Lawrence Berkeley National Laboratory}

\begin{abstract}
Gravitational lensing due to the large-scale distribution of matter in the cosmos distorts the primordial Cosmic Microwave Background (CMB) and thereby induces new, small-scale \textit{B}-mode polarization. This signal carries detailed information about the distribution
of all the gravitating matter between the observer and CMB last scattering surface. We report 
the first direct evidence for polarization lensing based on purely CMB information, from using the four-point correlations of even- and odd-parity \textit{E}- and \textit{B}-mode polarization
mapped over $\sim30$ square degrees of the sky measured by the \pb\ experiment. These data were
analyzed using a blind analysis framework and checked for spurious systematic contamination using
null tests and simulations. Evidence for the signal of polarization lensing and lensing \textit{B}-modes is found at
4.2$\sigma$ (stat.+sys.) significance. The amplitude of matter fluctuations is measured with a precision of $27\%$, and is found to be consistent with the Lambda Cold Dark Matter ($\Lambda$CDM) cosmological model.
This measurement demonstrates a new technique,
capable of mapping all gravitating matter in the Universe, sensitive to the sum of neutrino masses,
and essential for cleaning the lensing \textit{B}-mode signal in searches for primordial gravitational waves. 
\end{abstract}

\maketitle

\textit{Introduction:}
As Cosmic Microwave Background (CMB) photons traverse the Universe, their paths are  gravitationally deflected by large-scale structures.  
By measuring the resulting changes in the statistical properties of the CMB anisotropies, maps of this gravitational lensing deflection, which traces large-scale structure, can be reconstructed.
Gravitational lensing of the CMB has been detected in the CMB temperature anisotropy in several ways: 
in the smoothing of the acoustic peaks of the temperature power spectrum~\cite{Reichardt:2008ay,Das:2010ga,Keisler:2011aw}, 
in cross-correlations with tracers of the large-scale matter distribution~\cite{Hirata:2004rp,Smith:2007rg,Hirata:2008cb,Bleem:2012gm, feng12, actcross, planckCIB}, 
and in the four-point correlation function of CMB temperature maps~\cite{Das:2011ak,Feng:2011jx,vanEngelen:2012va, plancklensing}. 

The South Pole Telescope (SPT) collaboration recently reported a detection of lensed polarization using the cross-correlation between maps of CMB polarization and sub-mm maps of galaxies from \herschelspire{}~\cite{hanson2013}. 
A companion paper to this one has also shown the evidence of the CMB lensing-Cosmic Infrared Background cross-correlation results using \pb\ data~\cite{dgpaper}, finding good agreement with the SPT measurements. This cross-correlation is immune to several instrumental systematic effects but the cosmological interpretation of this measurement requires assumptions about the relation of sub-mm galaxies to the underlying mass distribution~\cite{snowmass}. 

In this Letter, we present the first direct evidence for gravitational lensing of the polarized CMB using data from the \pb\ experiment.
We present power spectra of the lensing deflection field for two four-point estimators using only CMB polarization data, and tests for spurious systematic contamination of these estimators. We combine the two estimators to increase the signal-to-noise of the lensing detection. 

\textit{CMB lensing:}
Gravitational lensing affects CMB polarization by deflecting photon trajectories from a direction on the sky ${\bf n}+{\bf d}({\bf n})$ to a new direction $\bf{n}$.  In the flat-sky approximation, this implies that the lensed and unlensed Stokes parameters are related by
\begin{equation}
(Q\pm iU)({\bf n})=(\tilde Q\pm i\tilde U)({\bf n}+{\bf d}({\bf n})), \label{plensed}
\end{equation}
where $\tilde{Q}$ or $\tilde{U}$ denotes a primordial Gaussian CMB polarization map, $Q$ and $U$ are the observed Stokes parameters, and ${\bf d}({\bf n})$ is the deflection angle. 
The CMB polarization fields defined in Eq.~(\ref{plensed}) are rotation-invariant under the transformation $e^{\pm 2i\phi}$ and can be decomposed into electric- ($\textit{E}$-) and magnetic-like ($\textit{B}$-) modes~\cite{Kamionkowski1997}.

Taylor expanding Eq.~(\ref{plensed}) to first order in the deflection angle reveals that the off-diagonal elements of the two-point correlation functions of $\textit{E}$- and $\textit{B}$-modes are proportional to the lensing deflection field, $d({\bf n})$.
Quadratic estimators take advantage of this feature to measure CMB lensing~\cite{Hu:2001tn, Hu:2001fa, HO}.
The two lensing quadratic estimators for CMB polarization are:
\begin{equation}
d_{EE}({\bf L})=\frac{A_{EE}(L)}{L}\int\frac{d^2{\bf l}}{(2\pi)^2}E({\bf l})E({\bf l'})\frac{C_l^{EE}{\bf L}\cdot {\bf l}}{\hat C_l^{EE}\hat C_{l'}^{EE}}\cos2\phi_{{\bf l}{\bf l'}},\label{EEest}
\end{equation}
and
\begin{equation}
d_{EB}({\bf L})=\frac{A_{EB}(L)}{L}\int\frac{d^2{\bf l}}{(2\pi)^2}E({\bf l})B({\bf l'})\frac{ C_l^{EE}{\bf L}\cdot {\bf l}}{\hat C_l^{EE}\hat C_{l'}^{BB}}\sin2\phi_{{\bf l}{\bf l'}}\label{EBest}.
\end{equation}
In Eqs.\ (\ref{EEest}, \ref{EBest}), ${\bf l}$, ${\bf l'}$, and ${\bf L}$ are coordinates in Fourier space with  ${\bf L} = {\bf l}+{\bf l'}$. 
The angular separation between ${\bf l}$ and ${\bf l'}$ is $\phi_{{\bf l}{\bf l'}}$, $C_l^{EE}$ is the theoretical lensed power spectrum, $\hat C_l^{EE}$ and $\hat C_{l}^{BB}$ are lensed power spectra with experimental noise. The estimators are normalized by $A_{EE}(L)$ and $A_{EB}(L)$ so that they recover  the input deflection power spectrum~\cite{HO}. 
 
The power spectrum of these estimators is:
 \begin{eqnarray}
 \langle d_{\alpha}({\bf L})d^{\ast}_{\beta}({\bf L'})\rangle&=&(2\pi)^2\delta({\bf L}-{\bf L'})  (C_L^{dd}+N^{(0)}_{\alpha\beta}(L)\\&+&\mbox{higher-order terms}).\nonumber 
 \end{eqnarray}
Here, $C_L^{dd}$ is the deflection power spectrum and $N^{(0)}_{\alpha\beta}$ is the lensing reconstruction noise, $\alpha$ and $\beta$ are chosen from $\left \{ \textit{EE}, \textit{EB} \right \}$, however we do not use $\alpha=\beta=EE$ as our focus is on the direct probe of CMB lensing represented by the conversion of \textit{E}-to-\textit{B} patterns. The $BB$ estimator also probes {\textit B}-modes, but it does not make a substantial contribution to the deflection power spectrum~\cite{HO}, so it is not used in this work. The four-point correlation function takes advantage of the fact that gravitational
lensing converts
Gaussian primary anisotropy to a non-Gaussian lensed anisotropy.  When
calculating this non-Gaussian signal, however, there is a ``Gaussian bias" term
$N^{(0)}$ which is the disconnected part in the four-point correlation that has to be subtracted.
The Gaussian bias is zero when $\alpha \ne \beta$ (i.e.,~$\langle d_{{\rm EE}}({\bf L})d^{\ast}_{\rm {EB}}({\bf L'})\rangle$) because $\langle E({\bf l})B({\bf l'})\rangle$=0 under the assumption of parity invariance. However, the Gaussian bias is much larger than the lensing power spectrum in the $\alpha=\beta$ case. 
The Gaussian bias can be estimated, and removed, in several ways \cite{Das:2011ak, vanEngelen:2012va, plancklensing};
the methodology employed in this Letter is described in the Data Analysis section.

\textit{Data Analysis:}\label{da}

The \Pb\ experiment \cite{Kermish_SPIE2012} is located at the James Ax Observatory in Northern Chile on Cerro Toco at West longitude $67^{\circ}47'10.4''$, South latitude $22^{\circ}57' 29.0''$, elevation 5.20 km.
The 1,274 polarization-sensitive transition-edge sensor bolometers are sensitive to a spectral band centered at 148 GHz with 26\% fractional bandwidth \cite{Arnold_SPIE2012}. 
The 3.5 meter aperture of the telescope primary mirror produces a beam with a 3.5$^\prime$ full width at half maximum (FWHM).
Three approximately $3^{\circ}\times 3^{\circ}$ fields centered at right ascension and declination (23h02m, $-32.8^{\circ}$), (11h53m, $-0.5^{\circ}$), (4h40.2m, $-45.0^{\circ}$), referred to as ``RA23", ``RA12", and ``RA4.5", were observed between May 2012 and June 2013. The patch locations are chosen to optimize a combination of low dust contrast, availability throughout the day, and overlap with other observations for cross-correlation studies.
  
The time-ordered data are filtered and binned into sky maps with $2'$ pixels. Observations of the same pixel are combined using their inverse-noise-variance weight estimated from the time-ordered data. All power spectra are calculated following the MASTER method~\cite{master}. 
We construct an apodization window from a smoothed inverse variance weight map. Pixels with an apodization window value below 1\% of the peak value are set to zero, as are pixels within $3'$ of sources in the Australia Telescope 20 GHz Survey \cite{AT20G}.
\textit{Q} and \textit{U} maps are transformed to \textit{E} and \textit{B} maps using the pure-B transform ~\cite{PureEstimator_Smith2006}.

We reconstruct the lensing deflection field by applying the two estimators in Eqs.~(\ref{EEest}, \ref{EBest}) to the sky maps for $l,l^\prime \in \{500,2700\}$. In these estimators, $C_l^{EE}, C_l^{BB}$ are calculated using \camb{} \citep{camb} for the WMAP-9 best-fit cosmological model. 
The theoretical deflection power spectrum, which is used in simulations, is estimated with \camb{} as well. 
We calculate power spectra for these reconstructions with the requirement that \textit{B}-mode information is included, 
thus there are two estimates of the lensing power spectrum: $\langle d_{EE} d^{\ast}_{EB} \rangle$ and $\langle d_{EB} d^{\ast}_{EB} \rangle$, hereafter referred to as $\langle EEEB\rangle$ and $\langle EBEB\rangle$ respectively. Intuitively, these two four-point correlation functions can be split into a product of two two-point correlations, $EE$ or $EB$, each of which is proportional to a deflection field (dark matter distribution) on the sky. So these four-point correlation functions estimate the squared deflection field which is proportional to the deflection power spectrum.
The first estimator $\langle EEEB\rangle$, which we will refer to as the cross-lensing estimator, is nearly free of Gaussian bias. 
The second estimator, $\langle EBEB\rangle$, requires calculation and removal of the large Gaussian bias~\cite{Feng:2011jx, Das:2011ak,vanEngelen:2012va, plancklensing}.   
The unbiased, reconstructed lensing power spectrum is calculated as follows: 
\begin{equation}
C_L^{dd}=(\langle d({\bf L})d^{\ast}({\bf L})\rangle-N_L^{(0)}) / T_L,
\end{equation}
where both the Gaussian bias $N_L^{(0)}$ and the transfer function $T_L$ are calculated using simulations. The mean estimated deflection is subtracted from the reconstructions and the realization-dependent Gaussian bias is subtracted for our final results.

We create 500 simulated lensed and unlensed maps to estimate the Gaussian bias and establish the lensing transfer function. 
The lensed and unlensed simulations are used in calculations to estimate the lensing amplitude and to test the null hypothesis of
no lensing, respectively. In the following context, ``lensed" or ``unlensed" refers to the case with or without lensing sample variance. We create map realizations of the theoretical spectra calculated by \camb.  In the lensed case, map pixels are displaced following Eq. (\ref{plensed}) to obtain lensed maps. We convolve each realization by the measured beam profile and filter transfer function, and add noise based on the observed noise levels in the polarization maps. 

We estimate the Gaussian bias by estimating the lensing power spectrum from a suite of unlensed simulated maps. 
The finite area of the \Pb{} fields results in a window function that couples to large-scale modes, biasing them at $l <300$. 
This low-$l$ bias has also been seen in temperature lensing reconstructions~\cite{planckT, biashardened}. 
After verifying with simulations that it is proportional to the lensing power spectrum, we correct this bias by calculating a transfer function derived from the ratio of the average simulated reconstructed lensing spectrum to the known input spectrum for $l<300$. 
This transfer function produces only $0.2\sigma$ difference in the overall significance of the two lensing estimators $\langle EEEB\rangle$ and $\langle EBEB\rangle$.  We validate the lensing reconstruction by correlating the estimated deflection fields from lensed map realizations with the known input deflection field. All the spectra for all patches and estimators agree with the input lensing power spectra.

\begin{figure*}
\rotatebox{-90}{\includegraphics[width=4cm, height=12cm, trim=7cm 0 0 0]{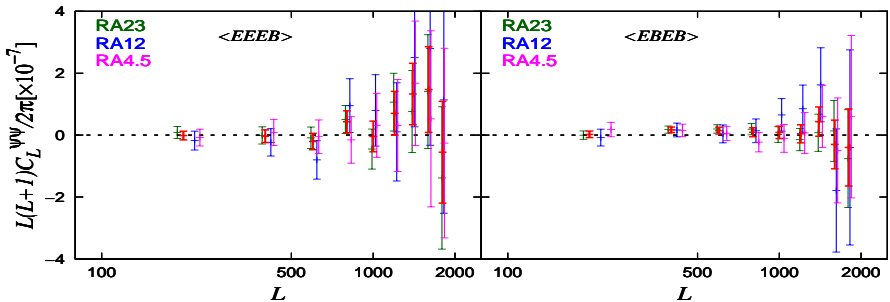}}
\caption{Curl null power spectra for each of the three patches for the $\langle EEEB\rangle$ and $\langle EBEB\rangle$ estimators. The patch-combined curl null power spectra are shown in red for the two lensing estimators. All the curl null power spectra are consistent with zero.}\label{curl}
\end{figure*}

\textit{Correlations between lensing estimators:}
Assuming CMB polarization is lensed, the two lensing estimators $\langle EEEB\rangle$ and $\langle EBEB\rangle$ make a correlated measurement of the lensing power spectrum. Monte Carlo simulations can precisely estimate these correlations~\cite{cov}. 
We produce 500 simulated lensing reconstructions for each lensing estimator, for each patch, and this correlation information is used to combine the two lensing estimators. 

The covariance matrix between two band-powers is defined as
\begin{equation}
\mathbf{C}_{AB} = \langle(C^{\textrm{sim}}_A -
\bar{C}^{\textrm{sim}}_A) (C^{\textrm{sim}}_{B} -
\bar{C}^{\textrm{sim}}_{B})\rangle;
\end{equation}
here the combined band-power is $C_A=(C_{\rm{channel{\ }1}}, C_{\rm{channel{\ }2}}, ...)$ and each $C_{\rm{channel{\ }X}}$ is co-added from simulations of all patches, with \rm{channel{\ }X either being $\langle EEEB\rangle$ or $\langle EBEB\rangle$ and $A$ or $B$  being the index of the band-power. 
The lensing amplitude ${\cal A}$ is constructed as 
\begin{equation}
{\cal A} = \frac{\sum_{AB} C^{(\textrm{th})}_A
\mathbf{C}^{-1}_{AB} C^{(\textrm{obs})}_{B}}
{\sum_{AB}C^{(\textrm{th})}_A\mathbf{C}^{-1}_{AB}C^{(\textrm{th})}_{B}}\label{CL}
\end{equation}
using \pb\ data (obs) and the WMAP-9 best-fit \lcdm\ model (th).
The variance of ${\cal A}$ is
\begin{equation}
(\Delta{\cal A})^2 = \frac{1}{\sum_{AB}C^{(\textrm{th})}_A
\mathbf{C}^{-1}_{AB} C^{(\textrm{th})}_{B}} \label{CLerror},
\end{equation}
and the significance of the lensing detection is ${\cal A}/\Delta{\cal A}$.\\

\textit{Estimation of systematic uncertainties:}
Systematic effects can generate spurious signals which could mimic the ones we want to probe. The statistical uncertainty of our measurements, which are $\Delta {\cal A}=0.30(0.47)$ for the unlensed (lensed) results, would overestimate the significance of our measurement if these systematic effects are neglected. We simulate the effect of measured instrument non-idealities and check the data for internal consistency and evidence of systematic instrumental errors using null tests. 
Leakage from temperature to polarization is constrained to be less than 0.5\% by correlating temperature maps with polarization maps.
A 0.5\% leakage from temperature-to-polarization in maps was simulated and found to introduce an error of $\Delta{\cal A}=\pm0.10 (\pm0.13)$ into the unlensed (lensed) simulations.
Polarized foregrounds are estimated based on models from the South Pole Telescope \cite{sptCl} assuming $5\%$ polarization fraction and constant polarization angle ~\cite{cmbpol_polarized_ps}.
This contamination was simulated and found not to bias the lensing estimators but it does increase the variance by an amount of $\Delta{\cal A}=\pm0.08 (\pm0.14)$ in unlensed (lensed) simulations.

We analyzed calibration and beam model uncertainty using lensed simulations. The beam model uncertainty is estimated from uncertainty in the point-source-derived beam-smoothing correction, and the variation in that correction across each field. 
We used the 1$\sigma$-bounds of the beam model as a simulated beam error and found that this created a change $\Delta{\cal A}={}^{+0.19}_{-0.16}$.
Absolute calibration error exists due to sample variance in the calibration to \lcdm\ (4\% including beam uncertainties), uncertainty in the pixel polarization efficiency (4\% upper bound), and uncertainty in the analysis transfer function (4\% upper bound), where all uncertainties are quoted in terms of their effect on $C_{\ell}^{BB}$ since these are conservative limits for error on $C_{\ell}^{EB}$ and $C_{\ell}^{EE}$~\cite{POLARBEAR2014}. 
We take 10\% as a bound on the calibration uncertainty, this corresponds to a calibration uncertainty of $\Delta{\cal A}={}^{+0.22}_{-0.18}$ in $\cal A$.
The total systematic error is $\Delta{\cal A}=\pm 0.13({}^{+0.35}_{-0.31})$ for unlensed (lensed) simulations. 

Null tests specific to the four-point lensing estimators are also examined. 
Deflection fields for different patches should be uncorrelated and this is used to test the lensing signals for potential contamination.
We define a ``swap-patch" lensing power spectra $C_L^{dd, \rm{null}}=\langle d_{{\rm patch}{\ }1}({\bf L})d_{\rm{patch}{\ }2}^{\ast}({\bf L})\rangle$ to test for contamination common to different patches~\cite{Das:2011ak}.
The deflection vector field can be decomposed into both gradient and curl components, of which only the gradient component is sourced by gravitational lensing (to leading order).
The curl power spectrum $C_L^{\psi\psi}$'s consistency with zero is thus another check of data robustness~\cite{curlnull}. 
While instrumental 
systematics could, in principle, mimic a lensing-like remapping of the CMB, such effects are  generically expected to produce both gradient and 
curl-like deflections.
A measurement of $C_L^{\psi\psi}$ is thus a sensitive test for instrumental systematics. 
Curl estimators are constructed by replacing $\bf{L}\cdot\bf{l}$ by $\bf{L}\times \bf{l}$ in Eqs.\ (\ref{EEest}, \ref{EBest}). 
For each of the null power spectra tests, a $\chi^2$ statistic is calculated assuming a null (zero signal) model. 
The probabilities to exceed the observed $\chi^2$ values are consistent with a uniform distribution from zero to one; the lowest PTE out of 15 tests (which include 9 swap-patch null and 6 curl null tests) is 8\%. For the curl null tests, the results are shown in Fig. \ref{curl}.

As a further systematic check, parallel work shows that the mass distribution information seen from the lensing reconstructions in this work is strongly correlated with cosmic infrared background maps from the \herschel\ satellite~\cite{dgpaper}.

In this work, for deflection power spectrum calculations, we adopted a blind analysis framework, whereby deflection power spectra were not viewed until the data selection and the analysis pipeline were established using realistic instrumental noise properties. 

\begin{figure*}
\rotatebox{-90}{\includegraphics[width=4cm, height=12cm, trim=7cm 0 0 0]{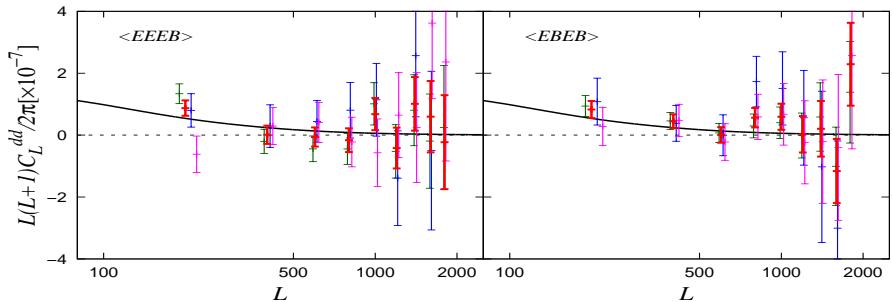}}
\caption{Measured polarization lensing power spectra for each of \pb 's three patches, for both lensing estimators $\langle EEEB\rangle$ (left) and $\langle EBEB\rangle$ (right). The lensing signal predicted by the $\Lambda$CDM model is shown as the solid black curve. The measured lensing power spectra are shown for each patch in dark green (RA23), blue (RA12) and magenta  (RA4.5), respectively and are offset in $L$ slightly for clarity. The patch-combined lensing power spectrum is shown in red.}\label{cldd_bmode_per_channel}
\end{figure*}

\begin{figure*}
\rotatebox{-90}{\includegraphics[width=4cm, height=12cm, trim=7cm 0 0 0]{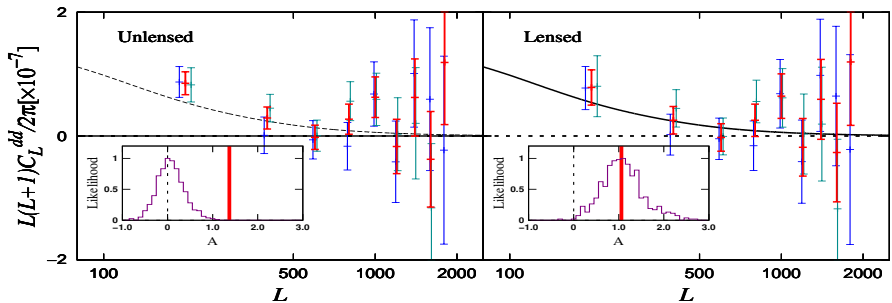}}
\caption{Polarization lensing power spectra co-added from the three patches and two estimators are shown in red.  
The lensing signal predicted by the $\Lambda$CDM model is shown as the dashed black curve in the left panel and the solid black curve in the right panel, respectively. 
The polarization lensing power spectrum $\langle EEEB\rangle$ is in blue and  $\langle EBEB\rangle$ dark green. \textit{Left:} A 4.2$\sigma$ rejection of the null hypothesis of no lensing. These data indicate a lensing amplitude ${\cal A}=1.37\pm 0.30\pm0.13$ normalized to the fiducial \lcdm\ value. \textit{Right:} The same data, assuming the existence of gravitational lensing to calculate error bars, including sample variance and including the covariance between $\langle EEEB\rangle$ and $\langle EBEB\rangle$. In this case, the lensing amplitude is measured as ${\cal A}=1.06\pm 0.47{}^{+0.35}_{-0.31}$, corresponding to 54\% uncertainty on the $C_L^{dd}$ power spectrum (27\% uncertainty on the amplitude of matter fluctuations). The histograms of the amplitudes ${\cal A}$ from 500 unlensed and lensed simulations are shown in the inset boxes. }\label{cldd_bmode}
\end{figure*}

\textit{Results:}
We present the polarization lensing power spectrum measurements for each of the three \pb\ patches and
the two $B$-mode estimators $\langle EEEB\rangle$ and $\langle EBEB\rangle$ in Fig.~\ref{cldd_bmode_per_channel}. The uncertainties in these band-powers do not include sample variance, that is, they represent the no lensing case. Fig.~\ref{cldd_bmode} shows the patches co-added, and the estimators $\langle EEEB\rangle$ and $\langle EBEB\rangle$ combined. The left panel does not assume the existence of lensing, and we measure a lensing amplitude of $1.37\pm 0.30\pm0.13$,
where the errors are statistical and systematic, respectively (this amplitude is normalized to the expected WMAP-9 \lcdm\ value). The rejection of the null hypothesis has a significance of 4.6$\sigma$ statistically and 4.2$\sigma$ combining statistical and systematic errors in quadrature. Without using $EE$ reconstruction to aid in the measurement of $E$-to-$B$ conversion, the lensing signal is detected at 3.2$\sigma$ significance statistically. 

The right panel of Fig.~\ref{cldd_bmode} assumes the predicted amount of gravitational lensing in the \lcdm\ model. 
In this case, the $\langle EEEB\rangle$ and $\langle EBEB\rangle$ estimators are correlated, which changes the optimal linear combination of the two, and requires that lensing sample variance be included in the band-power uncertainties.
Under this assumption, the amplitude of the polarization lensing power spectrum is measured to be ${\cal A}=1.06\pm 0.47{}^{+0.35}_{-0.31}$.  The last term gives an estimate of systematic error. Since ${\cal A}$ is a measure of power and depends quadratically on the amplitude of  the matter fluctuations, we measure the amplitude with 27\% error. The measured signal traces all the \textit{B}-modes at sub-degree scales. This signal is presumably due to gravitational lensing of CMB, because other possible sources, such as gravitational waves, polarization cosmic rotation~\cite{jonCB} and patchy reionization are expected to be small at these scales. 

\textit{Conclusions:}
We report the evidence for gravitational lensing, including the presence of lensing $B$-modes, directly from CMB polarization measurements. These measurements reject the absence of polarization lensing at a significance of  $4.2\sigma$. 
We have performed null tests and have simulated systematics errors using the measured properties of our instrument, and we find no significant contamination. Our measurements are 
in good agreement with predictions based on the combination of the \lcdm\ model and basic gravitational physics.
This work represents an early step in the characterization of CMB 
polarization lensing after the precise temperature lensing measurement from Planck. The novel technique of polarization lensing will allow future experiments to go beyond Planck in signal-to-noise and scientific returns. Future measurements will exploit this powerful 
cosmological probe to constrain neutrino masses~\cite{snowmass} and 
de-lens CMB observations in order to more precisely probe \textit{B}-modes 
from primordial gravitational waves.

\textit{Acknowledgments:}
This work was supported by the Director, Office of Science, Office of High Energy Physics, of the U.S. Department of Energy under Contract No. DE- AC02-05CH11231. The computational resources required for this work were accessed via the GlideinWMS~\cite{Sfiligoi2009} on the Open Science Grid~\cite{Pordes2008}. 
This project used the \camb{} and FFTW software packages. 
Calculations were performed on the Department of Energy Open Science Grid at the University of California, San Diego, the Central Computing System, owned and operated by the Computing Research Center at KEK, and the National Energy Research Scientific Computing Center, which is supported by the Department of Energy under Contract No. DE-AC02-05CH11231. 
The \Pb{} project is funded by the National Science Foundation under grant AST-0618398 and AST-1212230.
The KEK authors were supported by MEXT KAKENHI Grant Number 21111002, and acknowledge support from KEK Cryogenics Science Center. 
The McGill authors acknowledge funding from the Natural Sciences and Engineering Research Council and Canadian Institute for Advanced Research. We thank Marc Kamionkowski and Kim Griest for useful discussions and comments.
BDS acknowledges support from the Miller Institute for Basic Research in Science, NM acknowledges support from the NASA Postdoctoral Program, and KA acknowledges support from the Simons Foundation. MS gratefully acknowledges support from Joan and Irwin Jacobs.
All silicon wafer-based technology for \Pb{} was fabricated at the UC Berkeley Nanolab. 
We are indebted to our Chilean team members, Nolberto Oyarce and Jose Cortes. 
The James Ax Observatory operates in the Parque Astron\'{o}mico
Atacama in Northern Chile under the auspices of the Comisi\'{o}n Nacional de Investigaci\'{o}n Cient\'{i}fica y Tecnol\'{o}gica de Chile (CONICYT).
Finally, we would like to acknowledge the tremendous contributions by Huan Tran to the \pb\ project.

\bibliography{cldd_prl_rev}

\end{document}